\documentclass[10pt]{article}
\usepackage[explicit]{titlesec}
\usepackage{hyphenat}
\usepackage{ragged2e}
\RaggedRight

\usepackage{garamondx}
\usepackage[T1]{fontenc}
\usepackage{amsmath, amsthm}
\usepackage{graphicx}
\usepackage[utf8]{inputenc} 

\usepackage{soul}
\setul{.6pt}{.4pt}

\usepackage{geometry}
\usepackage{fancyhdr}
\usepackage[labelfont={footnotesize,bf} , textfont=footnotesize]{caption}
\makeatletter
\renewcommand\tagform@[1]{\maketag@@@ {\ignorespaces {\footnotesize{\textbf{Equation}}} #1.\unskip \@@italiccorr }}
\makeatother
\titleformat{\section}
  {\normalfont}{\thesection}{1em}{\MakeUppercase{\textbf{#1}}}
\titlespacing\section{0pt}{0pt}{-10pt}
\titleformat{\subsection}
  {\normalfont}{\thesubsection}{1em}{\textit{#1}}
\titlespacing\subsection{0pt}{0pt}{-8pt}

\makeatletter
\newcommand\sixteen{\@setfontsize\sixteen{17pt}{6}}
\renewcommand{\maketitle}{\bgroup\setlength{\parindent}{0pt}
\begin{flushleft}
\sixteen\bfseries \@title
\medskip
\end{flushleft}
\textit{\@author}
\egroup}
\makeatother

\usepackage[biblabel]{cite}
\title{Dynamic microscopic optical coherence tomography to visualize morphological and functional micro-anatomy of the airways}

\author{Tabea Kohlfaerber,$^{1,*}$ Mario Pieper,$^{2,3}$ Michael Münter,$^{4}$ Cornelia Holzhausen,$^{2}$ Martin Ahrens,$^{3,4}$ Christian Idel,$^{5}$ Karl-Ludwig Bruchhage,$^{5}$ Anke Leichtle,$^{5}$ Peter König,$^{2,3}$ Gereon Hüttmann,$^{1,2,3,6}$ and Hinnerk Schulz-Hildebrandt$^{1,2,3,**}$ \\ \\
$^{1}$Medizinisches Laserzentrum Lübeck GmbH,Peter-Monnik-Weg 4, 23562 Lübeck, Germany\\ 
$^{2}$University of Lübeck, Institute of Anatomy, Ratzeburger Allee 160, 23562 Lübeck, Germany\\
$^{3}$Airway Research Center North Member of the German Center for Lung Research, DZL, 22927 Großhansdorf, Germany\\
$^{4}$University of Lübeck, Institute of Biomedical Optics, Peter-Monnik-Weg 4, 23562 Lübeck, Germany\\  
$^{5}$University Medical Center Schleswig-Holstein, ENT Clinics, Ratzeburger Allee 160, 23538 Lübeck, Germany\\
$^{6}$Center of Brain, Behavior and Metabolism (CBBM), University of Lübeck, Ratzeburger Allee 160, 23562 Lübeck, Germany\\ \\
$^{*}$tabea.kohlfaerber@posteo.de \\
$^{**}$hinnerk.schulzhildebrandt@uni-luebeck.de
}

\begin{document}
\begin{justify}
% Makes the title and author information appear.
\vspace*{.01 in}
\maketitle
\vspace{.12 in}

% Abstracts are required.
\section*{abstract}
Imaging airway tissue, optical coherence tomography (OCT) provides cross-sectional images of tissue structures, shows cilia movement and mucus secretion, but does not provide sufficient contrast to differentiate individual cells. By using fast sequences of microscopic resolution OCT (mOCT) images, OCT can picture small signal fluctuations (dynamic) OCT and overcomes the lack in contrast caused by speckle noise. In this way, OCT visualizes airway morphology on a cellular level and allows to track the dynamic behavior of immune cells, as well as mucus transport and secretion.   
Here, we demonstrate that mOCT, by using temporal tissue fluctuation as contrast (dynamic OCT), provides the possibility to study physiological and pathological tissue processes \textit{in\,vivo}.

\section{Introduction}
 Microscopic evaluation of airway tissue is a long established method in diagnostic and research. Especially the assessment of surgical biopsies is a cornerstone in the diagnostic of airway diseases, but the invasiveness of the preceding excision poses a significant hurdle for a wide application. Furthermore, fixation, dehydration, and staining terminate any biochemical processes and only a static image is created. Therefore, biopsies can not investigate dynamic processes. Airway function, however, is characterized by fast processes such as ciliary beating and mucus transport and in comparison relatively slow processes such as immune cell trafficking. Therefore, techniques that allow imaging of the airways with microscopic resolution \textit{in\,situ} can be important tools for understanding the pathophysiology of airway diseases and enhance diagnostics. 

One option for imaging cellular processes \textit{in\,vivo} is two-photon laser scanning fluorescence microscopy \cite{Denk73}. In biomedical research it increased the understanding of immune processes immensely, but has its limitations. Unfortunately, dyes or fluorescent proteins  are usually required to provide specific contrast. However, fluorescent dyes can be toxic in high concentration and can bleach rapidly \cite{Germain2006}. In addition, only a very limited number of dyes are approved for the use in humans and these dyes have limited specificity for labeling specific structures. 
To avoid the disadvantage of exogenous dyes autofluorescence can be used. A study in the murine airways demonstrated that multiphoton microscopy (MPM) allows to investigate tissue morphology and cellular dynamics using only endogenous fluorophores \cite{Kretschmer2016}. However, not all tissue components can be visualized by autofluorescence. Cilia and mucus remain invisible. Furthermore, phototoxicity is a severely limiting factor in autofluorescence based MPM \cite{Klinger2015}.
Thus, an imaging system to quantitatively characterize the structure and functional dynamics of cells in living tissues without external dyes or photodamage is needed. 

Another optical imaging method is optical coherence tomography (OCT) which is based on reflected light and uses interference to reconstruct tomographic images of the tissue structures \cite{Fujimoto2000}. Due to its ability to detect structural changes and its low phototoxicity, OCT has become a standard technique in the fields of ophthalmology and dermatology \cite{Adhi2013,Kashani2017,Olsen2018}. While clinically used OCT does not achieve cellular resolution, the use of broadband light sources in combination with high numerical aperture optics pushes resolution to a few micrometers \cite{Liu2013, Pieper20}. OCT systems with microscopic resolution (mOCT) were already applied to living airway epithelium, organ cultures \cite{Liu19, Oldenburg2012}, excised tissues \cite{Oldenburg2012} and living animals \cite{Pieper20}, including measurements of ciliary beat frequency (CBF), glandular function, and mucus transport without exogenous labeling \cite{Liu2013}. Various studies were published that investigate key functional parameters, such as airway surface liquid and periciliary liquid depth, CBF, and mucociliary velocity of airways with microscopic resolution OCT \cite{Liu2013, Rehman2015, Leung2019, Pieper20}. However, the missing cell-specific contrast and speckle noise make it nearly impossible to investigate individual cells within the airway tissue.
Interferometric detection renders OCT sensitive to the phase of the scattered light. Dynamic tissue changes like blood flow, diffusion, or active motion of the scattering structures change the phase and as a result the intensity of the speckle pattern in the OCT images,  even if they are much smaller than the optical resolution. Measuring time series of B-scans over several seconds with 10-millisecond temporal resolution, the short time signal fluctuation (<~1\,s) can contrast cells and other tissue components \cite{Apelian2016,Muenter2020}. Detected are motions of cellular structures with nanometer sensitivity and 1\,µm spatial resolution. By Color-coding the frequency of the mOCT signal fluctuations, images reminiscent of histology can be obtained \cite{Muenter2020, Leung2020}.
This so-called dynamic mOCT (dmOCT) was first implemented using a time-domain full-field OCT (TD-FF-OCT) which generates high-resolution en face images \cite{Apelian2016}. Recently, we and others could demonstrate dynamic contrast with the more widely used scanning frequency-domain OCT (FD-OCT), which provides high-quality sectional images (B-scans) and high flexibility of choosing the image field \cite{Muenter2020, Leung2020}.

The aim of this work is to show the added value of dynamic contrast for OCT imaging of airways. We demonstrate dmOCT imaging of \textit{ex\,vivo} airway tissue samples of the murine trachea and the human nasal concha. Analysis of the short time signal fluctuations (<~1\,s) enabled clear discrimination of cells from their surrounding. Prolonged imaging allowed us to follow the migration and morphological changes of immune cells and mucus transport. In this sense, dmOCT provides high-resolution, high-contrast imaging of the airways over tens of minutes with a time resolution of milliseconds simultaneously. 

\section{Materials and methods}

\subsection{Setup and Processing}

A schematic principle of the mOCT setup is shown in Fig.\,\ref{fig:SetUp} and described in detail in \cite{Horstmann2017, Muenter2021}. To achieve isotropic spatial {1.5\,µm} resolution filtered light of a supercontinuum laser (SuperKExtreme EXW-OCT, NKT Photonics, Birkerød, Denmark) was used in combination with a 10x/0.3\,NA microscope objective (HCX APO L 10x/0.3 WUVI, Leica Microsystems, Wetzlar Germany). 
\begin{figure}[ht]
\centering
\includegraphics[width=0.6\textwidth]{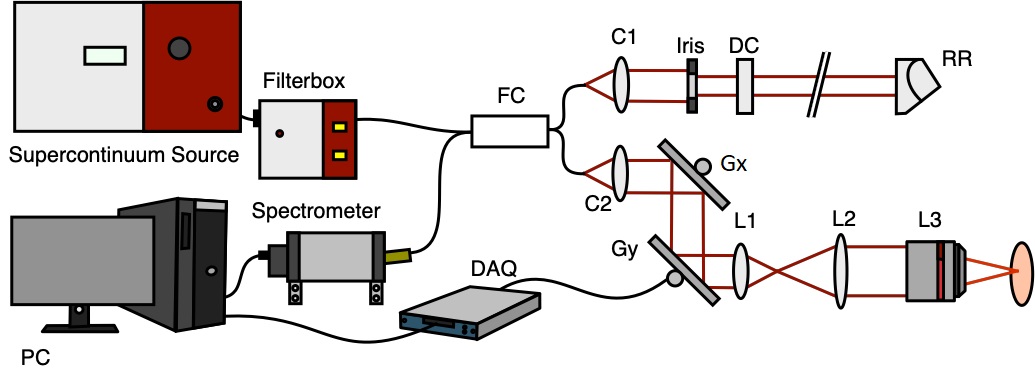}
\caption{(a) Schematic mOCT setup. FC: 50/50 fiber coupler; C1/C2: collimators; Gx, Gy: galvanometer mirror scanners; L1, L2: beam expander lenses; L3: microscope objective; DC: dispersion compensation; RR: retroreflector; DAQ, data acquisition device; PC: computer for data acquisition and scanning control.}
\label{fig:SetUp}
\end{figure}The light from the sample combined with the reflected light of the reference path results in an interference signal, which was measured by a customized high-speed spectrometer (Thorlabs GmbH, Bergkirchen, Germany), which covers a spectral range from 550\,nm to 950\,nm. Interference spectra were recorded by the CMOS camera with a maximum A-scan rate of 248\,kHz (OctoPlus CL, Teledyne, e2v, Canada). 
For dynamic contrast OCT imaging acquisition, 150 B-scans, which consisted of 500 A-scans, were used. A-scan rate was reduced to 100\,kHz for a better signal-to-noise ratio. Volumes were recorded by stacking a series of B-scans with increasing y-direction. With an effective B-scan rate of 111\,Hz and a total recording time of 1.35\,s, the frequency of signal fluctuations was measured between 0 to 55.5\,Hz \cite{Muenter2020}. The absolute values of the OCT signal in each pixel were Fourier transformed in time and the integral amplitude was calculated in three frequency bands. The three resulting values were color-coded in an RGB image, with blue representing slow-motion frequencies (<\,0.4\,Hz), green representing medium motion frequencies (0.4\,-\,4.9\,Hz), and red representing fast motion (4.9\,-\,25\,Hz). Each color channel was normalized from 0 to 1 and its histogram was adjusted to match the histogram of the image resulting from the log standard deviation of the OCT signal variation in each pixel. The images of the three  channels were median filtered afterwards. The computations were performed with Matlab (MATLAB R2019b, The MathWorks, Inc., U.S.).
To obtain dynamic contrast time lapse images, the acquisition of a B-scan with dynamic contrast was automated and could be performed within any time interval. Here, B-scans with dynamic contrast were recorded every 5 or 10 seconds. For automatic measurement  a custom build acquisition software using LabVIEW (National Instruments) was used.
This data processing resulted in false-color images and image series.

\subsection{Data evaluation and visualization}
For the measurement of the ciliary beat frequency (CBF), areas in the dmOCT images with cilia were isolated based on coloration. Color segmentation was done using the color thresholder implemented in the MATLAB software. After manually selecting the color values for cilia a binary segmentation mask was created for each dynamic image. This mask was used for cilia segmentation in all 150 B-scans of one data set. At each voxel of the chosen region of interest (ROI), the dominant frequency was determined by finding the frequency with the highest amplitude in a frequency range above 3\,Hz. High pass filtering avoided disturbance by the DC component and slowly varying frequencies. The dominant frequency was color coded and afterwards superimposed on the structural gray scale mOCT images. A cut-off amplitude was implemented for rejecting background noise at regions with low OCT signal. Computations were implemented with MATLAB.

Visualization, assembly of images series to videos and manual tracking of immune cells was done with Fiji (1.51n)\,\cite{Schindelin2012}.  

\subsection{Biological sample}
Experiments were performed on mouse trachea and human nasal concha \textit{ex\,vivo}. 

\subsection*{Murine tissue}
To obtain the murine trachea, the mice (C57BL/6 ) were sacrificed via inhalation of isoflurane, the trachea was extracted, cut open, placed in a petri dish with the epithelial facing upwards. The tissue was submerged in HEPES buffered Ringer solution for imaging and kept at approximately 25°C during experiments.
100\,µM of ATP-\(\gamma\)-S (Abcam, Berlin, GERMANY) was applied with a pipette to the tissue during measurements to increase ciliary beating frequency and induce mucus secretion and transport.

\subsection*{Human tissue}
Nasal concha mucosa was obtained from a patient who had undergone sinus surgery at the ENT department of University Medical Center Schleswig-Holstein (Lübeck, Germany). The sample was stored immediately in isotonic saline and directly investigated. All protocols were approved by the ethics committee of the University of Lübeck (16-278). All clinical examinations were carried out according to the principles of the Declaration of Helsinki (1964).

\subsection{Histology}
For tissue preservation, the samples were placed in 4\% paraformaldehyde after the experiments. Mouse trachea was embedded in epoxy resin to produce semi-thin sections of about 0.5\,µm thickness. To stain the semi-thin sections the dye methylene blue-azure\,II was used.
The human tissue of nasal concha was embedded in paraffin, stained with Hematoxylin-Eosin and cut into 4-5\,µm thick slices.

\section{Results}

\subsection{Dynamic contrast mOCT of murine airways}

\begin{figure}[ht]
\centering
\includegraphics[width=0.75\textwidth]{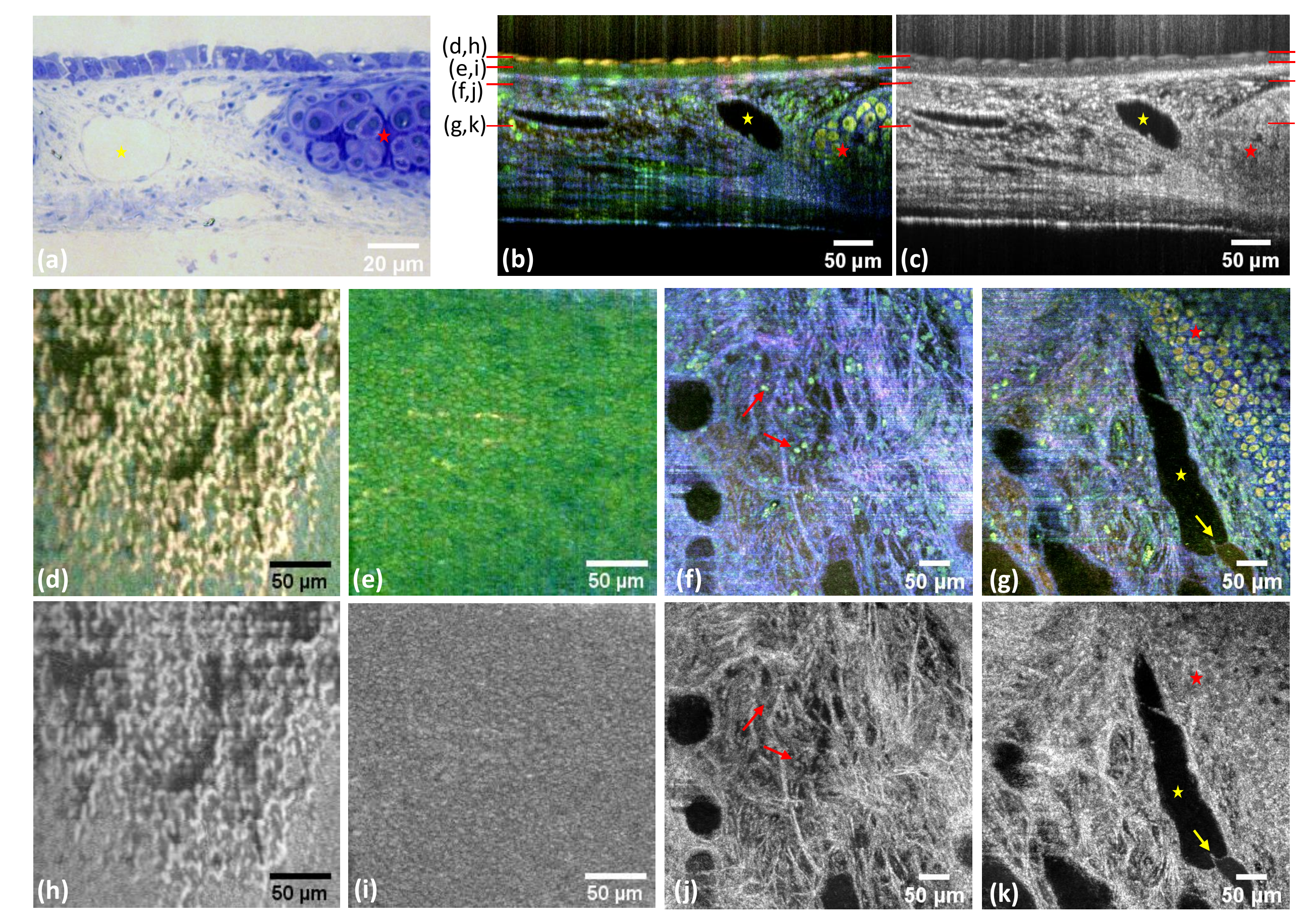}
\caption{Images of \textit{ex\,vivo} mouse trachea from one acquired volume. (a): exemplary semi thin section stained with methylene blue-azure II from a murine trachea. (b): dmOCT B-scan and (c): corresponding averaged mOCT B-scan. (d-k): \textit{en\,face} images of the same tissue at planes marked red in the preceding B-scans. Dynamic contrast is shown in the center line (d-g) and the corresponding averaged mOCT image below (h-k). (d,h): cilia, (e,i): epithelium, (f,j): connective tissue layer with immune cells (marked by red arrows), (g,k): connective tissue, cartilage (red star), lymphatic vessel (yellow star) with valve (marked by yellow arrow)}
\label{fig:ErgebnisseEnFace}
\end{figure}

With respect to epithelial structure and blood vessel supply, the mouse trachea is quite similar to a small human bronchus\cite{Kretschmer2016}. It served here as a model system for evaluating the visibility of cell structures in airways with dynamic contrast mOCT.

The trachea is characterized by different cellular and extracellular structures (Fig.\,\ref{fig:ErgebnisseEnFace}\,(a)). The innermost layer is a pseudostratified columnar epithelium, which contains mucus-producing goblet cells and ciliated cells which are responsible for mucus transport. Below appears fibrous tissue containing scattered fibroblasts, blood and lymphatic vessels, and mucus producing glands (not visible here). Within the submucosa cartilage rings with a characteristic pattern chondrocytes are embedded in a dense collagenous extracellular matrix.

By analyzing the frequency of the local signal fluctuations, these structures can be identified in unfixed, unstained living tissue (Fig.\,\ref{fig:ErgebnisseEnFace}\,(b,\,d-g)). Displaying the frequency content of the fluctuation spectra in the three color channels blue (constant signal and slow fluctuations), green (medium fluctuation frequencies), and red (fast fluctuations) a strong contrast between the different tissue structures is generated. 
 
Cilial beating causes strong signal fluctuation in all three spectral channels and appears as an intensive  yellow color. Ciliated cells are easily identified in cross-sectional and \textit{en\,face} images (Fig.\,\ref{fig:ErgebnisseEnFace}\,(b,\,d)). Averaging all recorded 150 B-scans, which are the basis of the frequency analysis (Fig.\,\ref{fig:ErgebnisseEnFace}\,(c)), cilia are also visible by the complete loss of speckle noise. 

Epithelial cell themselves appear in green color and are sharply delineated by color from the basal membrane and the submucosa. In the averaged mOCT B-scans the epithelium differs from the submucosa by a considerably lower OCT signal. Cellular structures are visible in \textit{en\,face} views of dynamic contrast imaging  (Fig.\,\ref{fig:ErgebnisseEnFace}\,(e)) as well as in averaged images (Fig.\,\ref{fig:ErgebnisseEnFace}\,(i)). However, contrast and delineation of the cells are better with dynamic contrast.
 In the submucosa, the lymph vessels appear as dark areas, marked by a yellow star in Fig.\,\ref{fig:ErgebnisseEnFace}\,(a-c,\,g,\,k). Cartilage, marked by a red star, can be identified with dynamic OCT by individual chondrocytes appearing in yellow color with a dark nucleus embedded in blue colored cartilage (Fig.\,\ref{fig:ErgebnisseEnFace}\,(b,g)). The averaged B-scans do not allow to identify individual chondrocytes. The cartilage appears as an area with a rather unspecific texture (Fig.\,\ref{fig:ErgebnisseEnFace}\,(c,k)). Fibers of the connective tissue give a strong signal in dynamic contrast (blue color) as well as in averaged mOCT images (Fig.\,\ref{fig:ErgebnisseEnFace}\,(b-c,\,f-g,\,j-k)). Individual cells within the connective tissue (red arrows) can be distinguished with certainty only by dynamic contrast and appear green. 

Dynamic contrast can be used to highlight individual cells and differentiate tissue structures in biological samples better than classical mOCT. 
In complex tissue types, such as airways, with many different cellular structures and highly dynamic processes such as ciliary beating, dmOCT provides significant advantages in cell phenotypes compared to mOCT. 

\subsection{Imaging dynamic processes by dmOCT}
\begin{figure}[ht]
\centering
\includegraphics[width=0.75\textwidth]{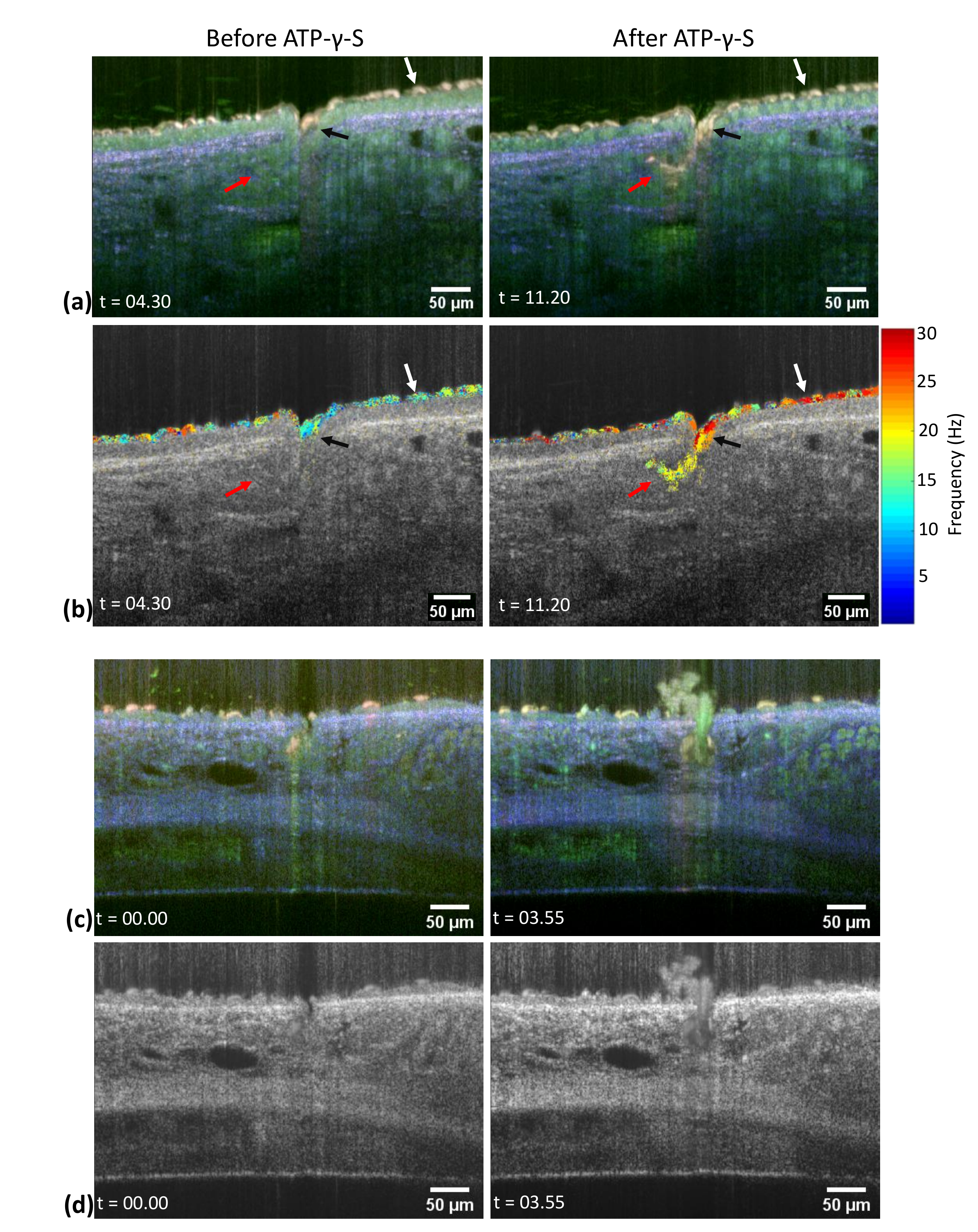}
\caption{Gland secretory duct in two \textit{ex\,vivo} mouse tracheas. B-scans respectively prior to and after ATP-\(\gamma\)-S stimulation. (a): dmOCT B-scans, (b): corresponding dominant frequencies of segmented cilia superimposed on the averaged gray scale image. Different cilia are indicated by a red, white and black arrow for further detailed frequency analysis. (c): dmOCT B-scan of a different sample during secretion (d): corresponding averaged mOCT B-scan. Entire time-lapse image sequence shown in supplementary videos\,i-iii. Time (t) is shown in minutes:seconds format.}

\label{fig:Gland_ATP_Freq}
\end{figure}
In the airways, different dynamic processes on different time scales are important for healthy function. The fastest process that repeats several times per second is ciliary beating, whereas mucus secretion and movement of cells of the immune system takes place over minutes or even hours. To image ciliary beating, images have to be acquired with high frame rates exceeding 50 frames per second. Imaging of mucus secretion requires imaging over several minutes and following the movement and interaction of cells of the immune system requires the acquisition of images ideally for more than an hour. Since the damage potential of mOCT is low\cite{Spicer2021} and imaging speed is high compared to other imaging technologies, mOCT provides optimal conditions observe these processes over several minutes or even hours. However, the missing cell contrast of OCT hampered its application so far.  

To assess the usability of dmOCT to analyze dynamic tissue processes that are taking place on very different time scales, B-scans with dynamic contrast were acquired for extended periods of time. Using this approach, we determined changes in ciliary beat frequency (CBF) following the application of ATP-\(\gamma\)-S. This stable analog of ATP activates extracellular receptors that result in increased CBF and mucus secretion from subepithelial glands. Using dmOCT we were able to detect increases in CBF in ciliated cells from gland ducts that are not easily identified on conventional mOCT images (Fig. \,\ref{fig:Gland_ATP_Freq}\,(a,b)). The color in the RGB image generated by dmOCT is not uniquely related to the CBF and therefore not suitable for quantitatively detecting changes in the CBF. Instead, the dominating frequency was calculated to determine the CBF and their change due to the application of ATP-\(\gamma\)-S in individual cells (Fig.\,\ref{fig:Gland_ATP_Freq}\,(a,b) and supplementary video i, ii). While Fig.\,\ref{fig:Gland_ATP_Freq}\,(a) shows B-scans prior ATP-\(\gamma\)-S administration (minute 4.30) and after (minute 11.20), (b) shows the same time points with the dominant CBF of the segmented cilia superimposed on the averaged mOCT B-scan. It can be seen that individual cilia have different dominant beating frequencies. Cilia on top of epithelium (white arrow in Fig.\,\ref{fig:Gland_ATP_Freq}\,(b)) starts around 15\,Hz and rises to 25\,Hz and cilia of the secretory duct of the gland (black arrow in Fig.\,\ref{fig:Gland_ATP_Freq}\,(b)) rises from an average 12\,Hz to 23\,Hz. while the cilia within the gland (red arrow) have a beating frequency between 14 and 20 Hz after ATP-\(\gamma\)-S application. Indicated by a red arrow in Fig.\,\ref{fig:Gland_ATP_Freq}\,(b) are cilia within the gland, which start to beat actively after ATP-\(\gamma\)-S application between 14-20\,Hz.

In addition to detecting changes in ciliated epithelial cells of the gland ducts, we also observed mucus secretion in response to ATP-\(\gamma\)-S (Fig.\,\ref{fig:Gland_ATP_Freq}\,(c,d) and supplementary video iii). Fig.\,\ref{fig:Gland_ATP_Freq}\,(c) illustrates a B-scan of the gland before and after mucus secretion
with dynamic contrast and Fig.\,\ref{fig:Gland_ATP_Freq}\,(d) shows the corresponding averaged mOCT B-scans. Secretion itself was readily observed by conventional mOCT \cite{Liu2013, Pieper20}, but imaging by dmOCT increased overall visibility of the mucus compared to the surrounding tissue. In addition, mucus color slightly changed during the secretion process indicating changes in the mucus during secretion (see supplement video\,iii). The underlying reason for this change remains to be determined but could be linked to mucus hydration during secretion \cite{Jaramillo2018}. 

\subsection*{Imaging immune cell trafficking}
\begin{figure}[t!]
\centering
\includegraphics[width=0.70\textwidth]{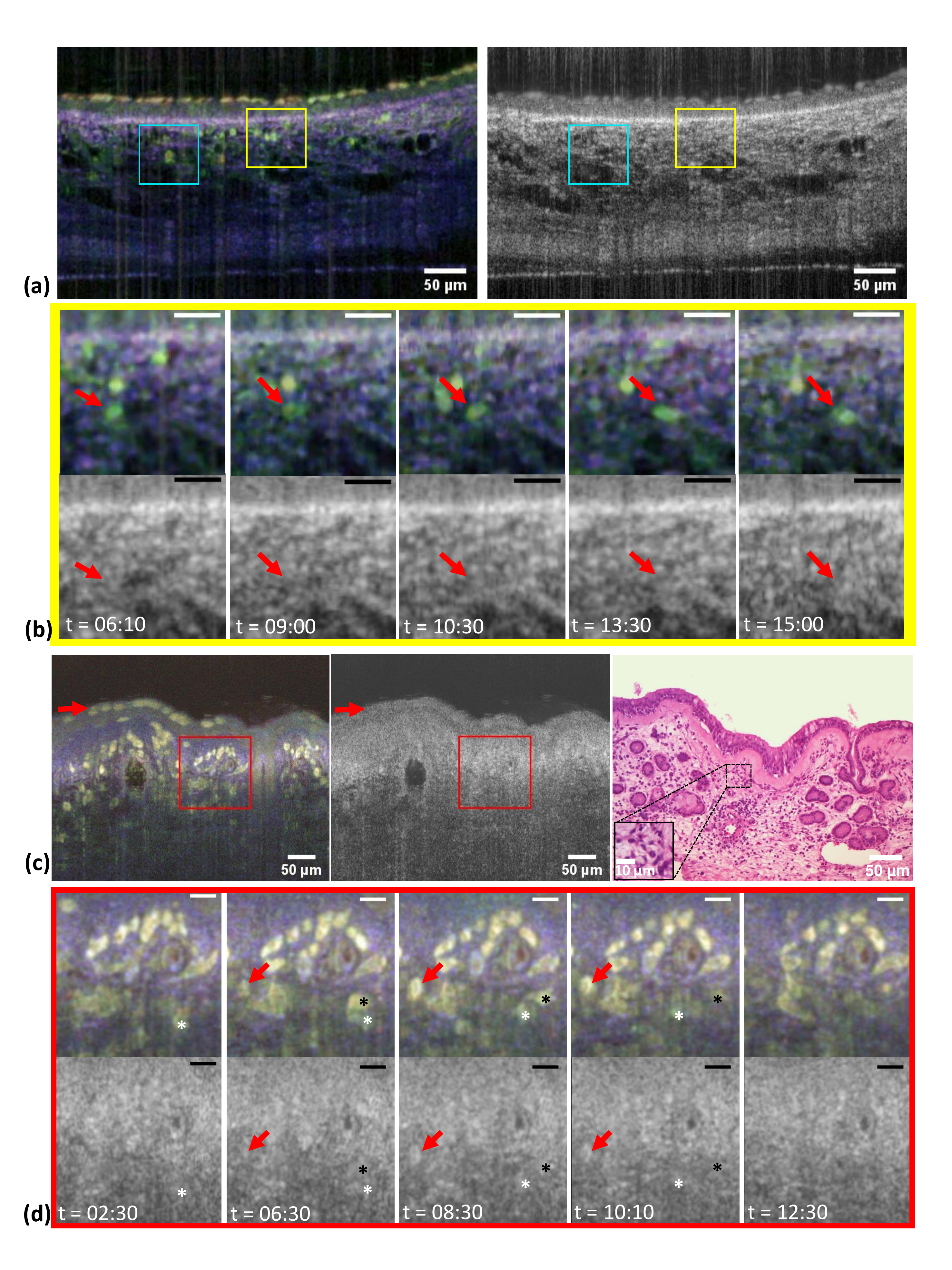}
\caption{Cell migration in airway tissue. (a) Cross-sectional dmOCT and averaged mOCT images of  \textit{ex\,vivo} mouse trachea (corresponding video\,iv  in supplement). (b): Series of images from the ROI indicated by the yellow square in (a) show the progression of an immune cell, indicated by red arrows at 5 different time points (mOCT on the top and dmOCT at the bottom). (c) Cross-sectional dmOCT, averaged standard mOCT image and H\&E stained histological section  of \textit{ex\,vivo} human nasal concha. Red arrow indicates immune cells in the epithelial layer (corresponding video\,v  in supplement). (d): series of images from ROI indicated in (c) by the red square shows an appearing and disappearing immune cell in the imaged B-scan layer (red arrow). The migration of two individual immune cells is indicated by a white and black star at 5 different time points (t). Scale bar (b,d) represents 20\,µm. Time (t) as shown in minutes:seconds.}
\label{fig:ImmuneCells}
\end{figure}
Using autofluorescence multiphoton microscopy we have previously shown that motile cells of the immune system are present in the connective tissue of the trachea \cite{Kretschmer2016}. To examine if dmOCT can detect the movement of mobile cells of the immune system we imaged for over 30\,minutes one dmOCT image every 10\,s. We were able to identify moving cells in the connective tissue below the epithelium which appeared in green color and whose movement could be followed over the entire time period (Fig.\,\ref{fig:ImmuneCells} (a,b) and supplement video\,vi). 
The time series of the magnified images in Fig.\,\ref{fig:ImmuneCells}\,(b) follows an individual immune cell (red arrow) at different time points. While the immune cell is clearly visible in the upper dynamic mOCT image in (b), it can not be detected in the averaged standard mOCT image below. The individual \textit{en\,face} images in (b) and the supplementary video iv show that the cell moves about 20\,\textmu m within 470\,s (7 minutes, 50 seconds (06:10-15:00)).
Since mOCT can be miniaturized to be used in endoscopes that are already used to examine the human nasal airways \cite{schulz2018, Cui2017, Gora2017}, dmOCT holds the potential to provide cell-specific contrast in endoscopic imaging \textit{in\,vivo}. 
As a first step to assess the potential of dmOCT for the analysis of nasal tissue, we examined excised specimen of human nasal concha which was removed during surgery. As in the murine trachea we were able to visualize subepithelial immune cells that were not visible in conventional mOCT. These cells also migrated through the tissue and could be followed over time (Fig.\,\ref{fig:ImmuneCells} (d) and supplement video\,v). Examination of nasal concha from several subjects showed variation in the number of inflammatory cells and epithelial integrity. In the example given in Fig.\,\ref{fig:ImmuneCells}\,(c-d) we detected a thickened basement membrane and inflammatory cells in the epithelial layer (red arrow in (c)).
The area marked by a red square in Fig.\,\ref{fig:ImmuneCells}\,(c) is magnified in (d) to show five images at different time points with dynamic contrast and averaged mOCT images.
The red arrows and the white and black asterisks each mark an immune cell moving through the connective tissue. 

Subsequent analysis of the specimen by H\&E histology confirmed the thickened basement membrane and the presence of various inflammatory cells in the connective tissue and the epithelium indicative of chronic inflammation such as lymphocytes and plasma cells. Although different cells of the immune system were present in the specimen, we did not see obvious color differences between individual cells, but a more specific analysis of frequency patterns might allow identification of individual cell types.  

Altogether, the results presented here show that time series of dynamic OCT images provide unique insights into complex biological processes of mucociliary transport or cell migration. The possibility of carrying out dynamic contrast imaging over a long period of time is further confirmed by the fact that no photodamage was observed in any of the time-lapse images. Our studies also indicate that the use of dmOCT could increase the information compared to conventional mOCT when examining human tissue \textit{in\,vivo}.

\section{Discussion}

The results of this study demonstrate the advantages of dynamic contrast compared to conventional mOCT imaging. mOCT achieves cellular resolution and can in principle image individual cells and subcellular structures. Nevertheless, cells within the tissue are often difficult to visualize due to speckle noise and lack of contrast. dmOCT relies on frequency analysis of intensity fluctuations at each pixel in one time series and provides the necessary contrast to reveal cells even within scattering tissue \cite{Muenter2020}. This allows a clear identification of individual cells within the connective tissue. Color is generated by assigning the mean signal intensity within a given frequency band to the colors blue, green, and red. The higher the mean signal, the brighter the color for each pixel. While in mOCT simple averaging only reduces speckle noise and nicely visualizes fibrous structures, single cells become visible by the addition of dynamic contrast. Epithelial cells, chondrocytes and immune cells exhibit cellular activity, which leads to a fluctuating signal at higher frequencies. Hence they appear in the green channel. 
Extracellular structures such as collagen fibers generate a constant signal and appear in the blue channel, which helps to identify these structures in dmOCT.
The fast movements of the cilia of the trachea generate a strong dynamic signal with a yellow color.
Interestingly, nuclei often appear black in both mOCT and dynamic OCT. Other studies with higher NA objectives and higher transverse resolution were able to recover some signal inside cell nuclei\cite{Thouvenin2017}. Thouvenin \textit{et\,al.} hypothesize that nuclei are appear somehow homogeneous at a resolution of 1\,µm and thus there is no backscatter and no dynamic signal from the nuclear volume \cite{Thouvenin2017}. 

Another technique that allows identification of metabolically active cells within airway tissue is multiphoton microscopy \cite{Kretschmer2013} which relies on the presence of endogenous fluorophores to visualize cells and second harmonic generation (SHG) to identify collagen fibers. Autofluorescence is mainly generated from NADH but other intracellular molecules are contributing to a lesser extent \cite{Croce2014}. NADH levels rely on cellular metabolism \cite{Szaszak2011} and consequently, NADH autofluorescence is dependent on cellular metabolism. 
A major drawback of autofluorescence multiphoton microscopy is that excitation of NADH requires wavelengths and laser intensities that can damage cells upon longer exposure \cite{Kretschmer2016} and makes this technique less suitable for prolonged imaging. The absence of photodamage in mOCT was recently investigated by Spicer\,\textit{et\,al.}\cite{Spicer2021} using a broadband (650-950\,nm) illumination as well. No photodamage to tracheal tissue was detected even after 60\,s of irradiation with 40\,mW laser power \cite{Spicer2021}. In comparison, in the positive control group stained with green tissue marking dye, laser-induced damage could already be detected after an exposure time of one second. Imaging for dynamic OCT is achieved in around one seconds and rapid scanning of the beam across the sample considerably reduces dwell time within a pixel. Hence, no laser induced damage can be expected and no cumulative effects in time lapse images were seen either. The safety of illumination enables dmOCT for an application in humans \textit{in\,vivo}.  
Additional drawbacks of conventional autofluorescence based multiphoton microscopy are that imaging is not fast enough to study processes such as ciliary beating and not all structures of interest exhibit endogenous fluorophores.
Cilia and mucus, two essential components of mucociliary clearance, are not visible in autofluorescence multiphoton microscopy. Both can readily be identified in mOCT due to their movement. By their continuous beating, cilia generate a highly dynamic signal, which is visible in dmOCT. Due to the integration of frequencies for the generation of dmOCT data, changes in ciliary beat frequency are not visible in dmOCT if the frequency changes only within one given frequency band. Nevertheless, the ciliary beat frequency can be calculated from the recorded data and can be determined for each ciliated cell. 

The increase in CBF at the epithelium after ATP-\(\gamma\)-S administration shown here  is consistent with the literature. After application of 5\,\(\mu\)M, 10\,\(\mu\)M, and 20\,\(\mu\)M ATP, CBF in mouse tracheal rings increased by $45\%$ \cite{Delmotte2006}, which is similar to our results. All measured values for CBF here prior ATP-\(\gamma\)-S stimulation are within the range expected for CBF by room temperature \cite{Delmotte2006,Liu2013,Rehman2015}.

Especially cilia in the gland secretory ducts are difficult to identify without dynamic contrast. 
Since no dominant CBF could be detected in cilia within the gland (Fig.\,\ref{fig:Gland_ATP_Freq}\,(b), red arrow) before ATP-\(\gamma\)-S application, it can be assumed that they do not beat actively until stimulated.

Interestingly, we also detected mucus that was freshly secreted from subepithelial glands. Even though mucus itself has no metabolism, its structure changes substantially during secretion due to unfolding and acquisition of water molecules \cite{Jaramillo2018} and it is possible that this process also generates a dynamic signal. Whether dmOCT can be used to specifically characterize mucus secretion is currently under investigation. 

In addition to ciliary beating and mucus secretion, we were able to track immune cells within \textit{ex\,vivo} airway tissue using time lapse imaging.
The locomotion of the cells and their morphology indicate positively that these are immune cells. The H\&E staining of the nasal concha tissue confirm the present of immune cells. It shows neutrophils, lymphocytes, plasma cells, possibly macrophages and eosiophils. However, it is currently not possible to differentiate by dmOCT between their various subgroups. Unfortunately, there are also no markers for OCT that can be used to label specific immune cells. 
In previous work \cite{Kretschmer2016} using two-photon microscopy and immunohistochemical staining to investigate the trachea and its immune cells, various immune cells were identified. MHCII positive cells were found in large numbers between the cartilage clamps throughout the connective tissue. Macrophages were also present in abundance throughout the connective tissue, although in smaller quantities than MHCII positive cells. T-cells were few throughout the connective tissue, and B-cells were even less likely to be found. Mast cells, on the other hand, were common throughout the connective tissue. In animals without acute airway inflammation, no or only occasional granulocytes were found.

The exact cellular mechanism underlying the fluctuations in signal intensity that are used for dmOCT is not yet understood. It is known that they rely on the metabolic activity of cells and it is reasonable to assume that movements of organelles within the cell or changes in the membrane reflectivity modulate the OCT signal \cite{Apelian2016, Scholler2020}. However, it is not known which organelles or what molecular or physical processes in the membrane are responsible. Insight into these mechanisms might allow identification of specific cell types based on their fluctuation spectrum and/or identification of the metabolic state of individual cells. 
At the moment, we cannot rule out that we miss cells in dmOCT because they do not  generate enough signal fluctuations. Further analysis combining dmOCT with fluorescence microscopy will help to clarify this important point.

Although dmOCT has unique advantages to study airways \textit{in\,vivo}, it has also limitations. Analysis of the dmOCT signal requires recording of images for about one second with minimal movement of the sample. This makes its use for \textit{in\,vivo} imaging difficult. However, OCT imaging has seen dramatic increases in speed during the last years which opens up the possibility to record 3D images at high frequencies. We recently presented an mOCT setup with a new high-speed spectrometer, which is capable of acquiring up to 600.000 A-scans/s and customized software allowed to image mOCT volumes with micrometer resolution at video rate\,\cite{Muenter2021}. High speed mOCT imaging in combination with  algorithms for motion correction shall allow the use of dmOCT \textit{in\,vivo}. 
In common with other label free techniques, dmOCT lacks the ability to identify specific cell types within the tissue which would be very helpful to assess the type of inflammatory cells in the tissue. It is tempting to speculate that deeper understanding of the underlying mechanisms that underlie dmOCT could lead to identification of signal components that could allow identification of individual cell types and/or their metabolic state. 

\section{Conclusion and Outlook}

The use of dmOCT opens up the possibility to study processes in the airways that have vastly different time scales ranging from ciliary beating over mucus secretion to the movement of immune cells. dmOCT is lable-free and, in contrast to fluorescence imaging, has no hazard of photochemical tissue damage. 

dmOCT is able to identify cells within the connective tissue only based the frequency analysis of intensity fluctuations of each pixel in one time series of about a second. It also visualizes the activation of cilia of the excretory duct of the gland. Their activity and the activity of cilia at the epithelium can be measured in parallel by calculation of the dominant frequency in the signal fluctuations. On a longer time scale (several minutes) the dynamic behavior of cilia function and mucus secretion is revealed. Over tens of minutes, dmOCT visualizes the dynamics of immune cells in the murine trachea and in freshly excised human nasal concha.
This could not be investigated with OCT until today, due to the missing contrast. 

Although we are not yet able to distinguish between different immune cells by dynamic contrast, the results shown here are the first step towards label-free investigation of immune processes using dmOCT. Hence dynamic contrast microscopic OCT combines important features of histology, OCT and MPM. To accurately determine the cells visible in the trachea by dmOCT, a direct comparison with other methods such as immunohistochemical staining is mandatory.

Without dynamic contrast, mOCT has already been incorporated in endoscopes with a frame rate  that exceeds video rate \cite{schulz2018, Cui2017, Gora2017}. If the speed of imaging can further be increased to correct for tissue movement, dynamic contrast could be an ideal tool to study airways in living animals and humans.
\end{justify}

\section*{Funding}
The work was supported by the German Ministry for Education and Research (82DZL001B2, 13N15846, 13N15847); European Union project within Interreg Deutschland-Denmark from the European Regional Development Fund (CELLTOM)

\section*{Acknowledgments}
We thank Kathy Budler from the Institute of Anatomy (Universität zu Lübeck) for preparing the H\&E slides and Jamin Koritke for the semi-thin sections. 

\section*{Disclosures}
The authors declare no conflicts of interest.

\section*{Visualization Availability}
The visualizations, are currently not publicly available, but can be obtained from the authors upon request.
\bibliographystyle{unsrt}

\bibliography{Ref}

\end{document}